%% file: 00_main.tex
\documentclass[conference]{IEEEtran}
\IEEEoverridecommandlockouts

\usepackage{cite}
\usepackage{hyperref}
\usepackage{amsmath,amssymb,amsfonts}
\usepackage{algorithmic}
\usepackage[pdftex]{graphicx}
\usepackage{subfigure}
\usepackage{textcomp}
\usepackage{booktabs}
\usepackage{flushend}
\usepackage{enumitem}
\usepackage{multirow}
\usepackage{siunitx}
\usepackage{url}
\usepackage[acronym,hyperfirst=false]{glossaries}
\usepackage[capitalise, noabbrev]{cleveref}

\def\BibTeX{{\rm B\kern-.05em{\sc i\kern-.025em b}\kern-.08em
    T\kern-.1667em\lower.7ex\hbox{E}\kern-.125emX}}

\newacronym{cnn}{CNN}{Convolutional Neural Network}
\newacronym{gan}{GAN}{Generative Adversarial Network}
\newacronym{tp}{TP}{True Positives}
\newacronym{fp}{FP}{False Positives}
\newacronym{tn}{TN}{True Negatives}
\newacronym{fn}{FN}{False Negatives}
\newacronym{tpr}{TPR}{True Positive Rate}
\newacronym{fpr}{FPR}{False Positive Rate}
\newacronym{dtw}{DTW}{Dynamic Time Warping}
\newacronym{vad}{VAD}{Voice Activity Detector}
\newacronym{tts}{TTS}{Text-to-speech}
\newacronym{vc}{VC}{Voice Conversion}
\newacronym{asv}{ASV}{Automatic Speaker Verification}
\newacronym{mos}{MOS}{Mean Opinion Score}
\newacronym{roc}{ROC}{Receiver Operating Characteristic}
\newacronym{auc}{AUC}{Area Under the Curve}
\newacronym{ir}{IR}{Impulse Response}
\newacronym{svm}{SVM}{Support Vector Machine}
\newacronym{gmm}{GMM}{Gaussian Mixture Model}
\newacronym{src}{SRC}{Sparse Representation Classifier}
\newacronym{dl}{DL}{Deep Learning}
\newacronym{mfcc}{MFCC}{Mel Frequency Cepstral Coefficient}
\newacronym{bfcc}{BFCC}{Bark Frequency Cepstral Coefficients}
\newacronym{lpcc}{LPCC}{Linear Prediction Cepstral Coefficients}
\newacronym{bed}{BED}{Band Energy Difference}
\newacronym{gsv}{GSV}{Gaussian Supervector}
\newacronym{nmse}{NMSE}{Normalized Mean-Squared Error}
\newacronym{ift}{IFT}{Inverse Fourier Transform}
\newacronym{rms}{RMS}{Root Mean Square}
\newacronym{snr}{SNR}{Signal-to-noise ratio}
\newacronym{awgn}{AWGN}{Additive White Gaussian Noise}
\newacronym{pcm}{PCM}{Pulse Code Modulation}
\newacronym{stft}{STFT}{Short-Time Fourier Transform}
\newacronym{nn}{NN}{Neural Network}
\def\BibTeX{{\rm B\kern-.05em{\sc i\kern-.025em b}\kern-.08em
    T\kern-.1667em\lower.7ex\hbox{E}\kern-.125emX}}
\begin{document}

\title{POLIPHONE: A Dataset for Smartphone Model Identification from Audio Recordings}
\author{\IEEEauthorblockN{
{Davide Salvi},
{Daniele Ugo Leonzio}, 
{Antonio Giganti},
{Claudio Eutizi}, \\
{Sara Mandelli},
{Paolo Bestagini},
{Stefano Tubaro}}
\IEEEauthorblockA{\textit{Dipartimento di Elettronica, Informazione e Bioingegneria (DEIB), Politecnico di Milano, Milan, Italy}}
}



\maketitle

\begin{abstract}
When dealing with multimedia data, source attribution is a key challenge from a forensic perspective. This task aims to determine how a given content was captured, providing valuable insights for various applications, including legal proceedings and integrity investigations. The source attribution problem has been addressed in different domains, from identifying the camera model used to capture specific photographs to detecting the synthetic speech generator or microphone model used to create or record given audio tracks.
Recent advancements in this area rely heavily on machine learning and data-driven techniques, which often outperform traditional signal processing-based methods.
 However, a drawback of these systems is their need for large volumes of training data, which must reflect the latest technological trends to produce accurate and reliable predictions.
This presents a significant challenge, as the rapid pace of technological progress makes it difficult to maintain datasets that are up-to-date with real-world conditions.
For instance, in the task of smartphone model identification from audio recordings, the available datasets are often outdated or acquired inconsistently, making it difficult to develop solutions that are valid beyond a research environment.
In this paper we present POLIPHONE, a dataset for smartphone model identification from audio recordings. It includes data from 20 recent smartphones recorded in a controlled environment to ensure reproducibility and scalability for future research.
The released tracks contain audio data from various domains (i.e., speech, music, environmental sounds), making the corpus versatile and applicable to a wide range of use cases.
We also present numerous experiments to benchmark the proposed dataset using a state-of-the-art classifier for smartphone model identification from audio recordings.
\end{abstract}


\input{01_introduction}

\input{02_background}
\input{03_dataset_generation}
\input{04_dataset_analysis}

\input{05_results}
\input{07_conclusion}

\section*{Acknowledgment}

We would like to sincerely acknowledge the editorial team of HWupgrade (\url{https://www.hwupgrade.it/}) for supplying us with the smartphones needed for the creation of this dataset.

This material is based on research sponsored by the Defense Advanced Research Projects Agency (DARPA) and the Air Force Research Laboratory (AFRL) under agreement number FA8750-20-2-1004. The U.S. Government is authorized to reproduce and distribute reprints for Governmental purposes notwithstanding any copyright notation thereon. The views and conclusions contained herein are those of the authors and should not be interpreted as necessarily representing the official policies or endorsements, either expressed or implied, of DARPA and AFRL or the U.S. Government.
This work was supported by the FOSTERER project, funded by the Italian Ministry of Education, University, and Research within the PRIN 2022 program.
This work was partially supported by the European Union under the Italian National Recovery and Resilience Plan (NRRP) of NextGenerationEU, partnership on ``Telecommunications of the Future'' (PE00000001 - program ``RESTART'', PE00000014 - program ``SERICS'').

{\small
\bibliographystyle{unsrt}
\bibliography{bibliography}
}

\end{document}

%% file: 01_introduction.tex
\section{Introduction}
\label{sec:introduction}

In the last few years, the creation and sharing of multimedia content have become highly accessible to the general public. This happened thanks to rapid technological advancements and the increasing availability of processing techniques and computational power.
As a result, the amount of media uploaded on the web and social platforms has been increasing daily, leading to the rise of new opportunities and techniques based on the use of the data itself.
At the same time, the creation of synthetic material and the alteration of existing media have become more accessible, with new technologies capable of generating fake data with impressive levels of realism~\cite{vice_dalle}.
This aspect is controversial, as misusing these powerful technologies could lead to unpleasant situations~\cite{MIT_lensa}.

Given the importance of multimedia content in today's society, it is becoming of paramount importance to determine the integrity of media under analysis and mitigate potential issues.
This is crucial not only to ensure people's cyber-security but also to use the media content as evidence in legal proceedings.
The scientific community is continuously working in this direction and proposing innovative solutions to address diverse multimedia forensic problems~\cite{berdich2023surveytransducer, verdoliva2020media}.
Among the considered tasks, one of the most crucial is source identification.
This consists of, given a multimedia signal, determining how it was generated or acquired.
The same problem has been addressed in several domains, ranging from images~\cite{mandelli2020cnn, cozzolino2019noiseprint, mandelli2022source}, videos~\cite{dal2021cnn, mandelli2019facing} to audio signals~\cite{salvi2022exploring, cuccovillo_acm_2022}.

In the case of audio signals,  numerous approaches have been proposed to address the challenge of identifying the recording microphone from a given track over the last few years.
The most recent solutions focus on using multi-class classifiers based on data-driven systems where each class represents a distinct microphone model~\cite{zeng2023source, zeng2024squeeze}.
While these methods have achieved excellent results, they are heavily reliant on the quality and scope of the training data. 
For a model to perform accurately and produce reliable predictions, it must be trained on large, diverse datasets.
Additionally, to ensure real-world applicability, the training data needs to incorporate cutting-edge technologies, as the rapid evolution of audio recording devices makes outdated data incomplete.
However, obtaining up-to-date and comprehensive datasets is a significant challenge, as many of the available datasets are either outdated, inconsistently collected or lack proper documentation, making them difficult to reproduce or apply in practical scenarios. 
This limitation impacts the state of research in this field, as the performance and accuracy of the developed classifiers depend significantly on the quality of the data used to train and test them.
Hence, the need for new data for identifying smartphone models from audio recordings is becoming critical.

In this paper we present and release POLIPHONE, a novel dataset for microphone model identification from audio recordings in the case of smartphones.
It contains recordings captured in a controlled environment by recent smartphones, bringing several contributions to advancing the state-of-the-art.
First of all, all the considered models are recent, as they have been released from \num{2018} onwards, making the dataset functional also for current real-world applications.
Secondly, the recording setup is entirely controlled, making the dataset reproducible and expandable with the addition of new microphone models in the future, even by other research groups. This is fundamental as the lack of a standard in the acquisition of recordings is one of the main reasons for the dearth of data in this field.
Finally, the recorded data comes from different audio domains, including clean and noisy speech, music and environmental sounds. This makes the data usable in different scenarios and for various applications.

The released dataset contains a total of more than \num{50} hours of audio recordings acquired with \num{20} different smartphones. The \glspl{ir} of each model is also provided, increasing the possible applications of the corpus.
Finally, we conduct a series of experiments to benchmark the proposed dataset and provide some information on the challenges it raises.
We do so by considering a state-of-the-art classifier presented in the literature, based on a \gls{cnn}~\cite{baldini_cnn_2019}.
Results confirm the need for new data to address the smartphone model identification task from audio recordings, as it could be an exciting topic for further research.

The rest of paper is structured as follows.
Section~\ref{sec:background} describes the background in the considered task, both regarding available data and presented methods. Here, we also provide a review of the most important datasets released in the field for research purposes, highlighting the motivations that led to the development of POLIPHONE.
Section~\ref{sec:generation} provides all the information about the presented dataset, illustrating the recorded corpus, the considered smartphones and the used recording setup.
Section~\ref{sec:analysis} presents the analysis performed on the released data, investigating the differences between the acquired recordings.
Section~\ref{sec:results} illustrates the experiments conducted to benchmark the released dataset, together with the achieved results.
Finally, Section~\ref{sec:conclusion} concludes the paper, states its main contributions, and briefly discusses possible future work.

%% file: 02_background.tex
\section{Background}
\label{sec:background}

This section provides the reader with background details regarding the problem of smartphone model identification from audio recordings.
First, we introduce the existing datasets in the state-of-the-art for the task at hand and outline their strengths and limitations.
This emphasizes the need for a novel set like the one we release in this paper.
Then, we provide an overview of the techniques presented in the literature to address the identification problem.
The information reported here can serve as a review of this research topic, which is often overlooked despite its importance in multimedia forensics.

\begin{table*}
\centering
\caption{Datasets proposed in the state-of-the-art for the microphone model identification task applied to smartphones.}
\label{tab:dataset_sota}
\resizebox{\textwidth}{!}{%
\begin{tabular}{@{}lccccccc@{}}
\toprule
Dataset Name & No. Brands & No. Models & Audio Type & No. Speakers & No. Env. & \begin{tabular}[c]{@{}c@{}}Recorded \\ Corpus\end{tabular} & Samp. Rate [\si{\kilo\hertz}] \\ \midrule
Buchholz et al.~\cite{buchholz_microphone_2009}, 2009 & 6 & 7 & Music, Noise, Pure Tone & N.A. & 12 & N.A. & 44.1 \\
Garcia-Romero et al.~\cite{garcia-romero_automatic_2010}, 2010 & N.A. & 8 & Speech & 61 & 1 & TIMIT & 8 \\
Hanilci et al. \cite{hanilci_brand_2012}, 2012 & 6 & 14 & Speech & 25 & 1 & TIMIT, various & 8 \\
Ikram et al.~\cite{ikram_higher_order_2012}, 2012 & 4 & 8 & Noise & - & 1 & - & N.A. \\
Cuccovillo et al.~\cite{cuccovillo_tampering_2013}, 2013 & 4 & 8 & Noise, Speech, Music & N.A. & N.A. & N.A. & 8 \\
Pandey et al.~\cite{pandey_psd_2014}, 2014 & 5 & 26 & Speech & N.A. & N.A. & N.A. & N.A. \\
Eskidere~\cite{eskidere_source_2014}, 2014 & 12 & 16 & Speech & 41 & 1 & TIMIT & 16 \\
Aggarwal et al.~\cite{aggarwal_cellphone_2014}, 2014 & 5 & 26 & Speech & N.A. & 1 & N.A. & N.A. \\
Kotropoulos et al.~\cite{kotropoulos_mobiphone_2014}, 2014 & 7 & 21 & Speech & 24 & 1 & TIMIT & 16 \\
Zeng J. et al.~\cite{zeng_21_2015}, 2015 & 13 & 21 & Noise & - & 1 & N.A. & N.A. \\
Zou et al.~\cite{zou_sparse_2017}, 2017 & 6 & 15 & Speech & 24 & 1 & TIMIT & 8 \\
Luo et al.~\cite{luo_bed_2018}, 2018 & N.A. & 31+141 & Speech & 4 & 4 & N.A. & 44.1 \\
Qin et al.~\cite{qin_cqt_2018}, 2018 & 7 & 24 & Speech & 172 & 1 & TIMIT, CKC-SD & N.A. \\
Verma et al.~\cite{verma_cnn_2019}, 2019 & 8 & 19 & Speech & 3 & 1 & N.A. & N.A. \\
Baldini et al.~\cite{baldini_cnn_2019}, 2019 & 4 & 32 & Pure Tones, Noise & - & N.A. & N.A. & 44.1 \\
Li et al.~\cite{li_gan_2019}, 2019 & 7 & 16 & Speech & 172 & 1 & TIMIT, various & N.A. \\
Lin et al.~\cite{lin_subband_2020}, 2020 & 5 & 20 & Speech, Various & N.A. & N.A. & N.A. & 16 \\
Zeng et al.~\cite{zeng_end_end_2020}, 2020 & 8 & 45 & Speech & N.A. & 1 & TIMIT & 32 \\
Shen et al.~\cite{shen_rars_2021}, 2021 & N.A. & 19 & Speech & 32 & N.A. & TIMIT & 16 \\
Hashim et al.~\cite{hashim_mobile_2021}, 2021 & 4 & 12 & Speech & 7 & N.A. & N.A. & 44.1 \\
Singhal et al.~\cite{singhal_wspire_2021}, 2021 & 5 & 5 & Whispered Speech, Speech, Pure Tones & 88 & 1 & MOCHA-TIMIT & 16 \\
Berdich et al.~\cite{berdich2022fingerprinting}, 2022 & 7 & 32 & Speech, Environmental Sounds & 24 & 4 & MOBIPHONE & 48 \\
\textbf{POLIPHONE, 2024} & \textbf{10} & \textbf{20} & \textbf{Speech, Music, Noise} & \textbf{30} & \textbf{1} & \textbf{VCTK, OrchSet, ESC-10} & \textbf{44.1} \\ \bottomrule
\end{tabular}%
}
\end{table*}

\subsection{Datasets for smartphone model identification from audio recordings}

When developing classification algorithms for recording smartphone model identification, one of the main requirements is to dispose of a large amount of data comprising audio recordings from several models.
Having data acquired from various sources will allow the classifier to learn how to differentiate between them.
For this reason, in recent years, the forensics community has released numerous datasets tailored for this purpose~\cite{ikram2012microphone, eskidere2014source, zeng2015audio}.
These have been developed considering different scenarios and application contexts and have been crucial in pushing research in the field.
Among these, a great effort has been invested in creating datasets containing tracks recorded by smartphones~\cite{hashim_mobile_2021, singhal_wspire_2021, berdich2022fingerprinting}. These models are the most prevalent at the consumer level and generate the vast majority of multimedia data that is shared online. Intending to make datasets that are practical for real-world applications, the inclusion of smartphones is a logical step.

While the existing datasets have been crucial in driving research in recording smartphone model identification, a comparison between them reveals several inconsistencies in data collection methods, which can hinder the performance of developed systems. The acquisition conditions vary considerably across recording setups and environments, resulting in a loose connection among most state-of-the-art corpora. 
This makes it difficult to compare the performance of the developed classifiers, which affects the reliability of the results, making them less applicable to real-world scenarios.
Additionally, as this research field is closely related to consumer devices, there is a need for continuous renewal of the data released to ensure that developed systems can also be usable in the real world.
For these reasons, the need for more data for the task of smartphone microphone model identification remains an ongoing challenge within the audio forensics community.

In this section, we collect the primary datasets presented in the literature for the task of smartphone model identification from audio recordings and highlight their strengths and weaknesses. 
We restrict our analysis to datasets that contain audio tracks recorded from smartphones.
\cref{tab:dataset_sota} lists them in chronological order, from the earliest to the most recent.
For further details regarding each corpus, we direct readers to the respective papers.

The majority of released datasets include only recorded speech tracks, while only a few consider audio data from different domains (e.g., music, noise, etc.)~\cite{buchholz_microphone_2009, cuccovillo_tampering_2013, berdich2022fingerprinting}.
This is presumably due to the fact that speech is the most acquired type of signal when dealing with smartphones, compared to others.
However, in a real-world scenario, recordings from other domains like environmental sounds, noise, and music may also be encountered, leading to the inclusion of other audio types in some sets.
Additionally, there are cases where pure tone signals are recorded to gather more information about the response of the microphones at specific frequencies~\cite{buchholz_microphone_2009, baldini_cnn_2019, singhal_wspire_2021}. 

Another critical aspect of the released data concerns the sampling rate of acquisitions. As it is shown in \cref{tab:dataset_sota}, most of the data were acquired at a relatively low frequency compared to the current recording standards employed by the newest models~\cite{garcia-romero_automatic_2010, hanilci_brand_2012, cuccovillo_tampering_2013, zou_sparse_2017}.
Although most of the speech content resides within the \num{0}-\SI{8}{\kilo\hertz} band, 
using a low sampling rate precludes studies regarding the impact of higher frequency bands on model identification problems.
This marks the need for data that aligns with the latest technological advancements, enabling the development of classification algorithms that can be more reliable in today's scenario.

Additionally, the majority of datasets presented in \cref{tab:dataset_sota} lack an explanation regarding the considered acquisition setup, both on the used equipment and the environmental parameters~\cite{pandey_psd_2014, lin_subband_2020, shen_rars_2021, hashim_mobile_2021}.
Many of them do not include technical specifications regarding recorded corpus, number of speakers, sampling rate, and bit depth.
This is a critical point regarding the reproducibility of the experiments, preventing a reliable comparison between different datasets and classification methods.
In some cases, highly specific scenarios are considered when performing the recordings. For instance, the authors of~\cite{berdich2022fingerprinting} consider car sounds recorded inside and outside the vehicles, making their experiments challenging to compare with those performed on other corpora.
Finally, most datasets reported here are not publicly available to the scientific community.
The only unrestricted sets are MOBIPHONE~\cite{kotropoulos_mobiphone_2014} and CCNU\_Mobile~\cite{zeng_end_end_2020}, which, however, include smartphones that are now outdated and not functional in the current state of the art.
This aspect further limits the research on the model identification task, making the proposed methods challenging to compare with each other.





\subsection{Methods for smartphone model identification from audio recordings}

In recent years, several methods have been proposed to address the recording smartphone model identification problem, leveraging different approaches~\cite{qin_cqt_2018, hashim_robust_2021, zeng2024spatio-temporal}.
Most of these are based on the analysis of acoustic features directly extracted from raw acquisition at the frame level.
This pipeline is usually implemented in two steps, where a feature extraction phase is followed by a supervised classification step.
Different acoustic features can be considered by these approaches, ranging from cepstral-based, spectral-based, and \emph{ad-hoc} features.
%
Among cepstral-based features, \glspl{mfcc} are the most adopted ones~\cite{hanilci_brand_2012, eskidere_source_2014, kotropoulos_mobiphone_2014, verma_recompressed_2018}, either directly used as input of the classifier or further processed before subsequent steps~\cite{zou_sparse_2015, li_spectral_clustering_2018}.
At the same time, spectral-based features are directly computed from the frequency representation of the signal and have proved their effectiveness in the model identification task in several studies~\cite{kotropoulos_source_2014, luo_bed_2018, cuccovillo_openset_2016, qin_cqt_2018}.
For instance, the authors of~\cite{berdich2022fingerprinting} explore the smartphone model identification task by employing the power spectrum of the recorded signals and feeding it to several supervised machine learning algorithms. 
Finally, problem-specific features have also been proposed.
These are not based on standard acoustic aspects but are designed to highlight some characteristics of the audio content that are helpful for the task at hand.
The authors of~\cite{jin_encoding_2019} propose features that exploit the information related to encoding, while others leverage different entropy measures to increase their robustness~\cite{baldini_entropy_2020, hashim_robust_2021}.

After the feature extraction phase, the classification step can be implemented using any supervised learning technique.
For instance, the authors of~\cite{jiang_kernel_2019, giganti2022speaker} use a \gls{svm} to classify non-linear features, while those of~\cite{hanilci_segment_2014} leverage \glspl{gmm}.
The first systems proposed for this task were based on classical machine learning approaches~\cite{hanilci_segment_2014, jiang_kernel_2019}.
However, most recent methods have applied \gls{dl} to the recording smartphone model identification tasks, developing powerful representations able to extract abstract and complex features from the input data.
In this context, \glspl{cnn} have been employed in~\cite{baldini_cnn_2019, verma_compressed_2021, naini_whispered_2022} on spectral-based features, while the authors of~\cite{lin_subband_2020} add an attention mechanism on a time-frequency representation of the recording, to emphasize the most informative frequency bands.
Deep autoencoder networks have been used to extract intrinsic signatures of the smartphones in~\cite{li_spectral_clustering_2018}, performing unsupervised clustering to aggregate recordings from the same microphone model. ResNet architecture has been used in~\cite{shen_rars_2021}, applying it to \glspl{mfcc} extracted from speech-free segments of the recording. The authors of~\cite{zeng_spatial_temporal_2021, zeng2024spatio-temporal} use a Bidirectional-LSTM architecture on cepstrum-based features sequentially extracted from the recording, considering the spatial and temporal information simultaneously. 
Finally, a \gls{nn} has been also applied in~\cite{giganti2022speaker, cuccovillo_acm_2022} to mitigate the effect of a noise-injection counter-forensic attack, comparing the benefits of this approach on three state-of-the-art cepstrum and spectral-based features.

%% file: 03_dataset_generation.tex
\section{Dataset Acquisition}
\label{sec:generation}

In this section, we provide all the information about POLIPHONE, the dataset we introduce and release in this paper.
First, we illustrate the structure of the recorded corpus.
Then, we present the smartphones and the recording setup we considered.
Finally, we describe the post-processing operations applied to the acquired tracks and detail the released data.

\subsection{Recording Setup}
\label{subsec:recording_setup}

One of the main aspects we considered when creating the POLIPHONE dataset was to ensure consistent acquisition conditions for all the smartphone models.
We did so for two main reasons. First, we wanted the differences between the tracks of the various smartphones to be solely due to their recording pipeline and not to external factors. This makes the audio data more trustworthy from a forensics point of view.
Second, fully controlling the acquisition conditions will allow us, in the future, to perform new recordings and expand the released dataset by integrating new models.
This is a crucial aspect in the model identification field as new smartphones are presented more and more frequently, and having a dataset that can be enlarged over time enables train classifiers on more comprehensive and updated data, leading to performances that are in line with the real-world technologies.

To achieve this goal, we organized the recording session in an anechoic environment, compliant with the ISO3745 standard~\cite{ISO3745}, playing back the tracks from a Genelec 8020C loudspeaker amplified by a Focusrite Scarlett 2i2 soundcard.
We ensured that the signals were stored in their raw form, capturing mono audio directly from the internal microphone of the models (the one at the bottom edge of the smartphone) without including any software post-processing or compression.
This approach was adopted to prevent the presence of additional software operations that could bias the recorded tracks.
The recordings were performed using the \textit{Auphonic} app, which is available for both Android and iOS devices.
All the data were acquired with a sampling frequency $f_\text{s} = $~\SI{44.1}{\kilo\hertz} and were saved in WAV \gls{pcm} encoded format at \SI{16}{\bit}.
We acquired the audio tracks considering \num{20} different smartphones, all released from late \num{2018} onwards.
\cref{tab:models} shows the complete list of the considered smartphone models.

\begin{figure}[t]
    \centering
    \includegraphics[width=.9\columnwidth]{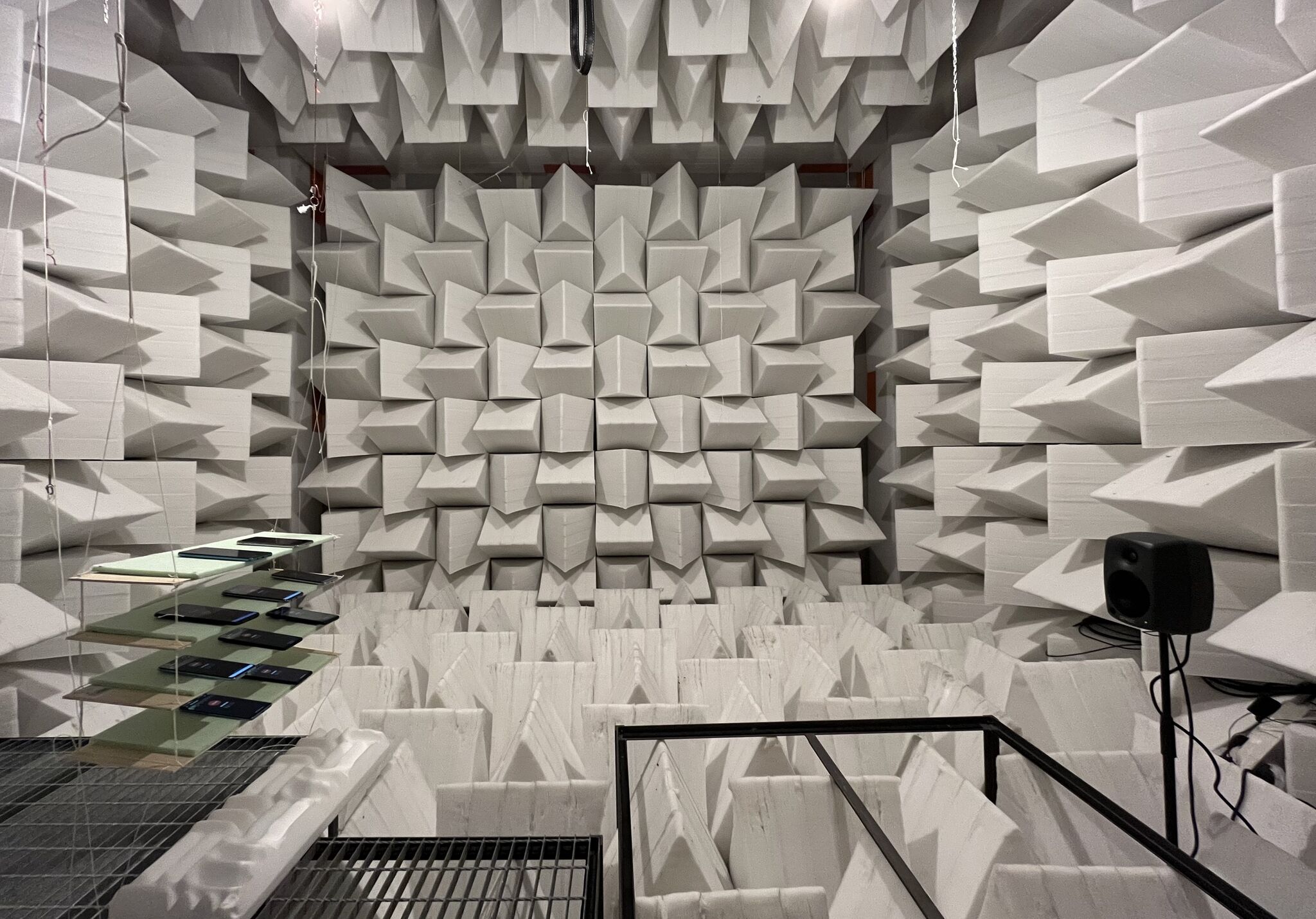}
    \caption{Recording setup used during the acquisition of the dataset tracks in the anechoic environment.}
    \label{fig:recording_setup}
\end{figure}

We structured the recording sessions in two batches, each involving \num{10} models, to guarantee that the acquisition conditions were the same for all smartphones and to eliminate differences due their positioning relative to the loudspeaker.
During the sessions, the phones were positioned approximately \SI{3}{\meter} away from the loudspeaker on a suspended structure to mitigate unwanted vibrations and ensure far-field recording conditions. \cref{fig:recording_setup} shows a picture of the considered recording setup.

\begin{table}[t]
\centering
\caption{List of the recorded smartphone models.}
\label{tab:models}
\begin{tabular}{cccc}
\toprule
No. & Model Name            & Brand    & Year \\ \midrule
1   & OnePlus Nord 2         & OnePlus  & 2021 \\
2   & Huawei P30             & Huawei   & 2019 \\
3   & Huawei Nova 9          & Huawei   & 2021 \\
4   & Huawei Nova 9 SE       & Huawei   & 2021 \\
5   & Motorola Edge 20       & Motorola & 2022 \\
6   & Motorola Edge 30       & Motorola & 2021 \\
7   & Motorola Moto G9 Power & Motorola & 2020 \\
8   & Realme GT              & Realme   & 2021 \\
9   & ROG Phone 3            & Asus     & 2021 \\
10  & ROG Phone 5            & Asus     & 2020 \\
11  & RedMagic 6             & RedMagic & 2021 \\
12  & Xiaomi 12              & Xiaomi   & 2021 \\
13  & Xiaomi 12X             & Xiaomi   & 2021 \\
14  & Xiaomi 12 Pro          & Xiaomi   & 2021 \\
15  & POCO M4 Pro            & POCO     & 2021 \\
16  & POCO X4 Pro            & POCO     & 2022 \\
17  & Redmi Note 11          & Redmi    & 2022 \\
18  & iPhone XS Max          & Apple    & 2018 \\
19  & iPhone 12 mini         & Apple    & 2020 \\
20  & iPhone 13              & Apple    & 2021 \\ \bottomrule 
\end{tabular}
\end{table}

\subsection{Recorded tracks}

The purpose of the proposed dataset is to provide a valuable corpus for the recording microphone model identification task in the case of smartphones.
As discussed in \cref{sec:background}, multiple studies have been carried out in this field, and most focus on signals containing speech.
Although speech signals are commonly acquired using models such as smartphones, testing the developed algorithms solely on such data is somehow reductive as it does not sample the generalization capabilities of the algorithms on different audio domains.
Also, mobile phone microphones are often optimized to work well on voice frequencies, as these are the most common ones they could encounter.
Analyzing the behavior of the microphones on full-spectrum signals could yield valuable insights for a more comprehensive analysis.
To make the released data as complete as possible, we recorded the audio tracks considering several scenarios, increasing the number of possible studies that can be performed.

The categories of audio signals that we recorded are: (i) speech, (ii) music, and (iii) environmental sounds, resulting in almost \num{150} minutes of audio data for each smartphone.
Additionally, we recorded a chirp signal, which can be used to extract the \gls{ir} of the single microphones and perform more in-depth studies.

Regarding speech, we considered the audio tracks from the VCTK corpus~\cite{yamagishi2019cstr} and we recorded data from \num{30} different speakers, with each speaker contributing for approximately \num{1} minute of recordings. 
We acquired the first \num{13} tracks of the corpus for each speaker, ensuring uniformity in the spoken sentences across all of them.
\cref{tab:speakers} shows a list of considered speakers.
Our selection aimed to encompass a scenario as broad as possible, i.e., balancing the number of male and female speakers and considering several different English accents.

In the POLIPHONE dataset, we release a single track for each speaker containing all the \num{13} utterances mentioned above.
At the beginning of each released track, we recorded \SI{5}{\second} of silence, which can be helpful in computing the microphone noise floor, as requested in some state-of-the-art studies. 
The total length of the recorded speech signals is approx. \num{40} minutes.
Additionally, we include a \textit{txt} file containing the text transcriptions of the considered utterances.
We recorded the speech signals in both clean and noisy conditions, injecting noise into the audio tracks before their acquisition.
This is paramount as this kind of noise is difficult to recreate through post-processing operations. However, it is a compelling use case to analyze, as it may influence some of the factors leveraged by the detectors to generate their predictions.
In particular, we recorded the same data of the clean speech case by injecting \gls{awgn} with a \gls{snr} of \SI{30}{\decibel}.

\begin{table}[t]
    \caption{List of the considered speakers from the VCTK corpus.}
    \label{tab:speakers}
    \begin{minipage}{.49\columnwidth}
    \centering
    \resizebox{\textwidth}{!}{
        \begin{tabular}{cccc}
        \toprule
        No. & VCTK ID & Gender & Accent         \\
        \midrule
        1 & P258    & M      & English        \\
        2 & P256    & M      & English        \\
        3 & P232    & M      & English        \\
        4 & P243    & M      & English        \\
        5 & P254    & M      & English        \\
        6 & P311    & M      & American       \\
        7 & P334    & M      & American       \\
        8 & P345    & M      & American       \\
        9 & P360    & M      & American       \\
        10 & P316    & M      & Canadian       \\
        11 & P347    & M      & South African  \\
        12 & P304    & M      & Northern Irish \\
        13 & P245    & M      & Irish          \\
        14 & P237    & M      & Scottish       \\
        15 & P251    & M      & Indian         \\ \bottomrule
        \end{tabular}
        }
    \end{minipage}
    \hfil
    \begin{minipage}{.49\columnwidth}
    \centering
    \resizebox{\textwidth}{!}{
        \begin{tabular}{cccc}
        \toprule
        No. & VCTK ID & Gender & Accent         \\
        \midrule
        1 & P225    & F      & English        \\
        2 & P228    & F      & English        \\
        3 & P229    & F      & English        \\
        4 & P239    & F      & English        \\
        5 & P240    & F      & English        \\
        6 & P294    & F      & American       \\
        7 & P297    & F      & American       \\
        8 & P299    & F      & American       \\
        9 & P300    & F      & American       \\
        10 & P303    & F      & Canadian       \\
        11 & P314    & F      & South African  \\
        12 & P238    & F      & Northern Irish \\
        13 & P266    & F      & Irish          \\
        14 & P262    & F      & Scottish       \\
        15 & P248    & F      & Indian         \\ \bottomrule
        \end{tabular}
        }
    \end{minipage}
    
\end{table}

As for the music tracks, we recorded the entire ORCHSET dataset~\cite{bosch2016evaluation}, which includes \num{64} audio excerpts of symphonic music, for a total of approximately \num{23} minutes.

Finally, for environmental sounds, we utilized the ESC dataset~\cite{piczak2015esc}, which is a collection of \num{5}-second-long tracks that include sound events.
In particular, we recorded the complete ESC-10 subset, which comprises \num{400} labeled environmental recordings (\num{10} classes, \num{40} clips per class), for a total of approximately \num{35} minutes of recorded signals.

For further details about these datasets, we refer the reader to the respective papers.
We release the recordings of both music and environmental tracks, adopting the same organization as the original sets to facilitate their use.
Since the frequency bands covered by these tracks are remarkably different from those of speech, their exploration in the recording model identification scenario could be fascinating.

As mentioned above, we also recorded a chirp signal with each smartphone.
We considered a sine sweep lasting \SI{5}{\second}, ranging from \num{20} to \SI{20000}{\hertz} with a sampling frequency $f_\text{s} = $~\SI{44.1}{\kilo\hertz}.
We can use these recordings to extract the \gls{ir} of each model's microphone.
The \glspl{ir} not only provide information about the microphones themselves but also allow to perform several experiments.
For instance, we can convolve a clean audio signal with the \gls{ir} of a specific smartphone, making the track sound as if it was recorded by that specific smartphone.

To compute the \gls{ir} of a microphone, we need to convolve its recorded sweep with the inverse of the original sweep~\cite{muller2001transfer}. This operation is performed in the frequency domain, where it corresponds to a simple multiplication.
After the convolution, we consider only the real part of the result and assume only the first half of its samples, which correspond to the linear \gls{ir} of the microphone.
Finally, the \gls{ir} is ready to use, allowing us to convolve it with a given audio track to produce the same signal, but as if it had been captured by the specified microphone.
The computation of the \glspl{ir} is possible thanks to the acquisition setup we considered.
In fact, given the controlled and anechoic conditions in which the tracks were recorded, we can consider the contribution of the recording environment as null and the impulse response as due solely to the model's microphone.

\subsection{Post-processing operations}
\label{sec:post-processing}

Although we considered the same acquisition setup across all the smartphones, the recorded tracks exhibit distinct characteristics among models.
In particular, both dynamic range and response at distinct frequencies are prone to high variations across different smartphones.
The cause of this phenomenon likely lies in the diverse hardware configurations and acquisition pipelines employed by individual smartphones, leading to easily distinguishable recorded signals (cf. \cref{sec:analysis}).
In order to make the recording smartphone model identification task more challenging within the POLIPHONE dataset, we aim to mitigate variations between tracks from different devices by conducting post-processing on the recorded data.

The post-processing pipeline we propose involves three steps.
First, we standardize the signals to have zero mean and unit standard deviation, ensuring that the signals are on a comparable scale.
Then, we multiply each signal by a \textit{tanh} function, aiming to reduce biases in the dynamic range of the signals~\cite{hermansky1994rasta}.
This operation is particularly beneficial when dealing with data exhibiting extreme values or widely varying magnitudes.
Finally, we resample the data from $f_\text{s}=$~\SI{44.1}{\kilo\hertz} to $f_\text{s}=$~\SI{16}{\kilo\hertz}.
While acknowledging the non-linear nature of the \textit{tanh} transformation, we believe that its use is necessary in our scenario. Specific models show recorded signals with notably different dynamic ranges compared to other devices, leading to highly distinguishable patterns.
By applying a \textit{tanh} transform, signal values are compressed towards the boundaries of the range $[-1, 1]$, aligning the dynamic ranges of the signals more closely. This adjustment significantly minimizes disparities between signals from different models and enhances overall uniformity in our dataset.

Formally, starting from a recorded signal $\mathbf{x}$, we compute its post-processed version $\mathbf{x}_\text{proc}$ as
\begin{equation}
\label{eq:post_processing}
	\mathbf{x}_\text{proc} = \tanh \left( \frac{\mathbf{x} - \mu(\mathbf{x})}{\sigma(\mathbf{x})} \right) ,
\end{equation}
where $\mu(\cdot)$ and $ \sigma(\cdot)$ are the mean and standard deviation operations, respectively.

\subsection{Released data}

The POLIPHONE dataset is open-source and can be downloaded at this link\footnote{\url{https://zenodo.org/records/13903412}}.
We release the data in three different versions to address various research and application needs. Here are the descriptions of each version:
\begin{enumerate}
    \item \textbf{original\_recordings}: This partition contains the unaltered, raw audio recordings as captured by the recording smartphone models. These files preserve the original quality and sampling frequency ($f_\text{s} = $~\SI{44.1}{\kilo\hertz}), making them ideal for those who require the highest fidelity for tasks such as detailed acoustic analysis or audio restoration.
    
    \item \textbf{normalized\_data}: These audio files have been processed following the pipeline described in \cref{sec:post-processing}. We release these tracks at both $f_\text{s} = $~\SI{44.1}{\kilo\hertz} and $f_\text{s} = $~\SI{16}{\kilo\hertz}.
    
    \item \textbf{convolved\_data}: This partition contains the clean audio tracks that have been convolved with the \glspl{ir} of specific smartphone models to simulate their acquisition by the microphone of those devices. This version is helpful for testing the robustness of the identification algorithms under more challenging multimedia forensic conditions.
    We release these tracks at $f_\text{s} = $~\SI{16}{\kilo\hertz}.
\end{enumerate}

In addition to these audio data, we release the chirp signals recorded from the smartphones and the corresponding \glspl{ir} we computed.
We also provide the code used to compute the \glspl{ir} from the recorded tracks in both Matlab and Python\footnote{
Code will be released upon acceptance.
}.

%% file: 04_dataset_analysis.tex
\section{Dataset Analysis}
\label{sec:analysis}

In this section, we analyze the recorded tracks, highlighting the dissimilarities between signals acquired by smartphones included in this dataset.
As detailed in \cref{subsec:recording_setup}, since the recording setup considered is consistent across all the tracks in the dataset, we can effectively compare them with each other and look for dissimilarities between different models.
All the differences identified between the tracks are attributable to distinct acquisition pipelines implemented by the models or the variations in their hardware.
In this case, we consider the original recordings acquired in the anechoic chamber at $f_\text{s} = $~\SI{44.1}{\kilo\hertz}, without applying any post-processing technique (i.e., version 1 of our dataset).

\begin{figure*}[t]
    \centering
    \includegraphics[width=\textwidth]{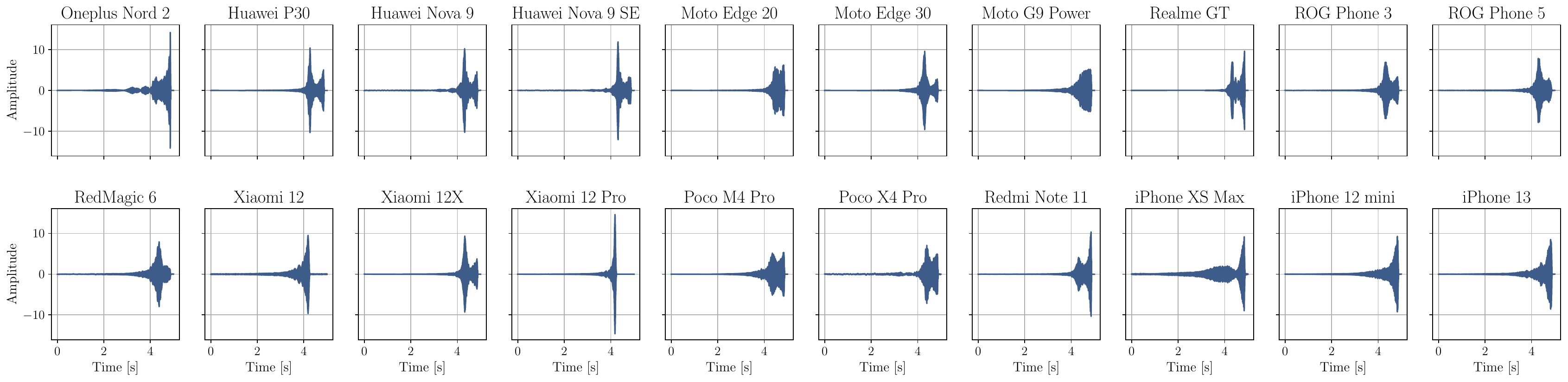}
    \caption{Waveforms of the sine sweep signals recorded by all the considered smartphone models.}
    \label{fig:sweeps}
\end{figure*}

\begin{figure*}[t]
    \centering
    \includegraphics[width=\textwidth]{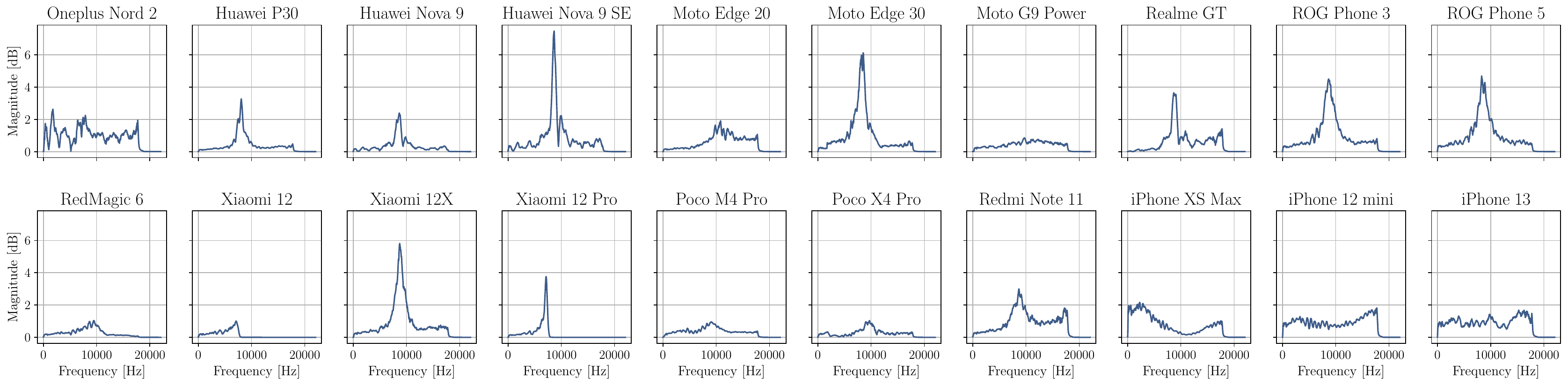}
    \caption{Spectral representations of the Impulse Responses of all the considered smartphone models.}
    \label{fig:IR_spectra}
\end{figure*}

\begin{figure*}[t]
    \centering
    \includegraphics[width=\textwidth]{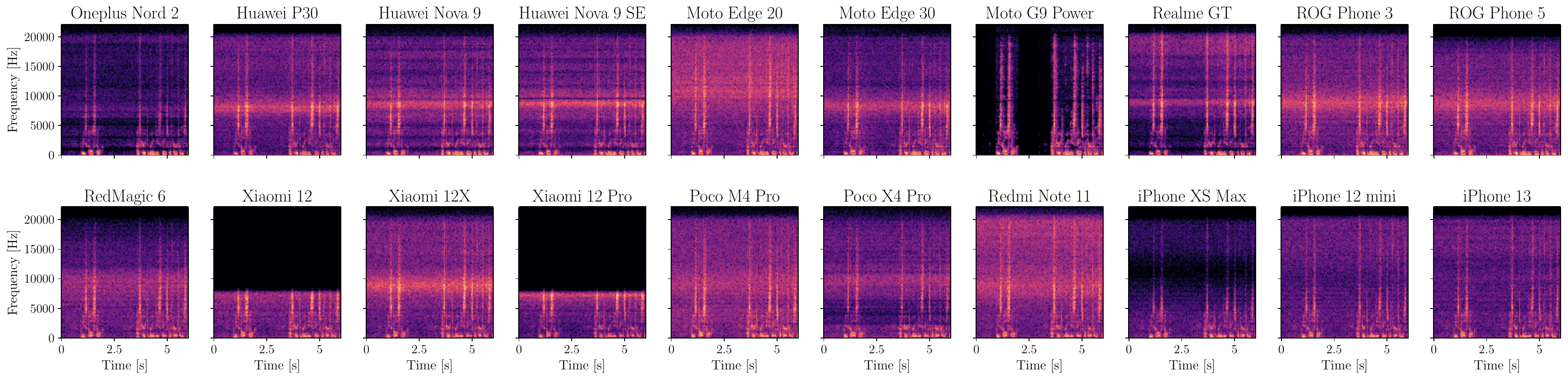}
    \caption{Mel-spectrogram representations of the noisy speech signals recorded by all the considered smartphone models.}
    \label{fig:noise_speech}
\end{figure*}

First, we focus our analysis on the recordings of sine sweep signals.
We do so since these recordings, although brief, include highly informative content regarding the microphone characteristics of the considered smartphones.
\cref{fig:sweeps} shows the waveforms of the recorded chirps, while \cref{fig:IR_spectra} shows the spectra of the \glspl{ir} extracted from them.

The acquired signals are very different from one model to another, with differences in both the dynamic range and the equalization at different frequencies.
We would expect an ideal acquisition to exhibit a monotonic increase in the signal amplitude over time and an almost flat spectrum for the extracted \gls{ir}, similar to those presented by models like iPhone 12 mini and Motorola G9 Power.
However, many of the devices do not present this behavior.
For instance, phones like Huawei P30 and ROG Phone 5 show an increased amplitude in the frequency bands that are typical of speech.
This behavior may be attributed to the design choices made by the manufacturers, which emphasize certain frequency bands to compensate for the limitations of the microphone hardware.
Indeed, smartphones are often provided with low-cost microphones, whose primary use is acquiring speech signals, whether it is during calls, voice messages, or video recordings. For this reason, it makes sense to emphasize certain frequency bands for better hearing results.

During the analysis, we noticed a particular behavior exhibited by two Xiaomi phones, namely the Xiaomi 12 and Xiaomi 12 Pro.
Despite their capability to capture audio data at \SI{44.1}{\kilo\hertz}, these models clip all the content above \SI{8}{\kilo\hertz}, as if acquiring at $f_\text{s} = $~\SI{16}{\kilo\hertz}.
This characteristic is even more evident by looking at \cref{fig:noise_speech}, which shows the Mel-spectrogram representations of the noisy speech signals recorded by all the considered smartphones.
In the case of these two models, a black horizontal band shows the lack of content above \SI{8}{\kilo\hertz}. Interestingly, this particular behavior is not present in the recordings of Xiaomi 12X, which is produced by the same manufacturer.

Another notable aspect observed in \cref{fig:noise_speech} belongs to Moto G9 Power.
As the recorded signal is injected with \gls{awgn}, we expect to observe content across the entire frequency range at each time instant.
However, it seems like this model includes a \gls{vad} mechanism and puts to zero all the time windows where no speech is detected. 
This behavior may prove advantageous or disadvantageous, depending on the purpose of the recording.
While it discards content of un-voiced segments, enhancing clarity in acquired speech, it also discards potentially valuable audio information.
As in the previous case, this behavior belongs only to this specific model and is not extended to other smartphones from the same manufacturer (i.e., Motorola Edge 20 and Motorola Edge 30).
We recall that we considered the same recording pipeline for all the models, meaning that these behaviors are likely due to hardware filtering or post-processing operations specific to each model.

%% file: 05_results.tex
\section{Smartphone Model Identification from Audio Recordings}
\label{sec:results}

In this section, we benchmark the released dataset on the task of microphone model identification in the case of smartphones. We do so by considering a method proposed in the literature for this specific task and testing it in various scenarios~\cite{baldini_cnn_2019}.
The goal of these experiments is to evaluate the contribution that POLIPHONE can make to state-of-the-art.

The problem we consider is a multi-class classification task in closed-set conditions, and it is formally defined as follows.
Let us consider a discrete-time input speech signal $\mathbf{x}$, sampled with sampling frequency $f_\text{s}$ and acquired by a smartphone belonging to a set of $I$ models $y_i \in \{y_0, y_1, ..., y_{I-1}\}$.
The goal of this task is to develop a classifier $\mathcal{C}$ able to predict which model $y_i$ has been used to record $\mathbf{x}$.

The considered classifier $\mathcal{C}$ is a \gls{cnn}-based model which takes as input the logarithmic \gls{stft} of the audio recordings under analysis~\cite{baldini_cnn_2019}.
For details regarding the model used, please refer to the original paper~\cite{baldini_cnn_2019}.

We trained the network for \num{100} epochs by monitoring the value of the validation loss.
We assumed \num{10} epochs as early stopping, a batch size of \num{32}, and a learning rate of \num{e-4} appropriately reduced on plateau.
We assumed Cross Entropy as loss function and RMSProp as optimizer.
The input of the network consisted of log-spectrograms derived from \SI{1}{\second} audio windows, with a configuration of $N_\text{FFT} = 2048$ points and non-overlapping Hanning windows. Each track of the dataset is represented by \num{15} windows, covering the initial \SI{15}{\second}. Longer tracks are trimmed, while shorter ones are replicated to match the desired length.



\subsection{Clean Speech experiments}
As a first experiment, we evaluate the considered baseline on the clean speech recordings of the POLIPHONE dataset.
We do so as this represents the most common scenario for the recording smartphone model identification task, and the baseline was specifically designed for this purpose.
Specifically, we consider the \emph{normalized\_data} partition of the dataset, which enhances the uniformity across all the classes (cf. \cref{sec:post-processing}).
We examine two different variants of this subset: the original release, which has a sampling frequency $f_\text{s} = $~\SI{44.1}{\kilo\hertz} and a resampled version at $f_\text{s} = $~\SI{16}{\kilo\hertz}.
This experiment aims to benchmark the dataset and to verify whether the most critical information for this task resides in the \num{0}-\SI{8}{\kilo\hertz} frequency band or in higher frequencies.

In this experiment, we divide the corpus into three subsets (train, validation, and test sets) using proportions of \num{60}\%, \num{20}\%, and \num{20}\%, respectively. In the case of speech data, we partition the dataset ensuring that speakers are disjointed across the three subsets.
We also train multiple iterations of the classifier, each time utilizing a distinct fraction of the training set, ranging from \num{10}\% to \num{100}\%.
This approach allows us to investigate how the classifier's performance evolves with varying access to training data.

\begin{figure}[t]
    \centering
    \includegraphics[width=.9\columnwidth]{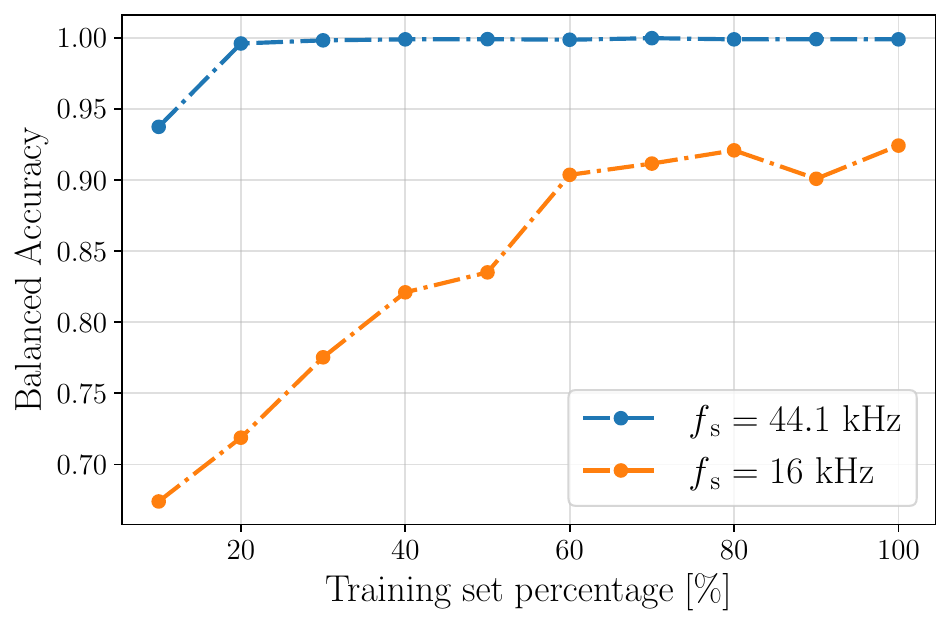}
    \caption{Balanced accuracy values of the considered baseline, trained on different percentages of the training set. 
    }
    \label{fig:baselines_percentage}
\end{figure}

\cref{fig:baselines_percentage} shows the results of this analysis considering balanced accuracy as a metric.
We observe a significant change in performance trends when considering different sampling frequencies.
The classifier trained and tested on \SI{44.1}{\kilo\hertz} data demonstrates nearly perfect accuracy, reaching almost 100\% in all scenarios except when trained on only 10\% of the training set, where performance drops by \num{5}\%. Conversely, the classifier that considers \SI{16}{\kilo\hertz} data achieves a maximum accuracy of \num{92}\% and experiences a substantial drop in performance as the training set size decreases. This indicates that the critical audio content for recording smartphone model identification is likely found in the frequency bands above \SI{8}{\kilo\hertz}, enabling the first classifier to perform exceptionally well even with limited training samples. In contrast, the second classifier's task is more challenging, necessitating more training data to enhance its classification performance.

\subsection{Different Domains' experiments}
We now broaden our investigations by exploring other scenarios beyond clean speech.
In this analysis, we train and test the considered classifier on audio data from diverse domains (i.e., clean speech, noisy speech, music, and environmental noises), investigating challenging cross-test situations.
As in our previous approach, the classifier is trained using a \num{60}\%-\num{20}\%-\num{20}\% split for train, validation, and test sets, and we consider the first \SI{15}{\second} of each audio track.
Based on the results of the previous experiment, we focus on the post-processed data at \SI{16}{\kilo\hertz} because the excessively high accuracy values achieved at \SI{44}{\kilo\hertz} may obscure the insights we gain from each experiment.

\begin{figure}[t]
    \centering
    \includegraphics[width=.8\columnwidth]{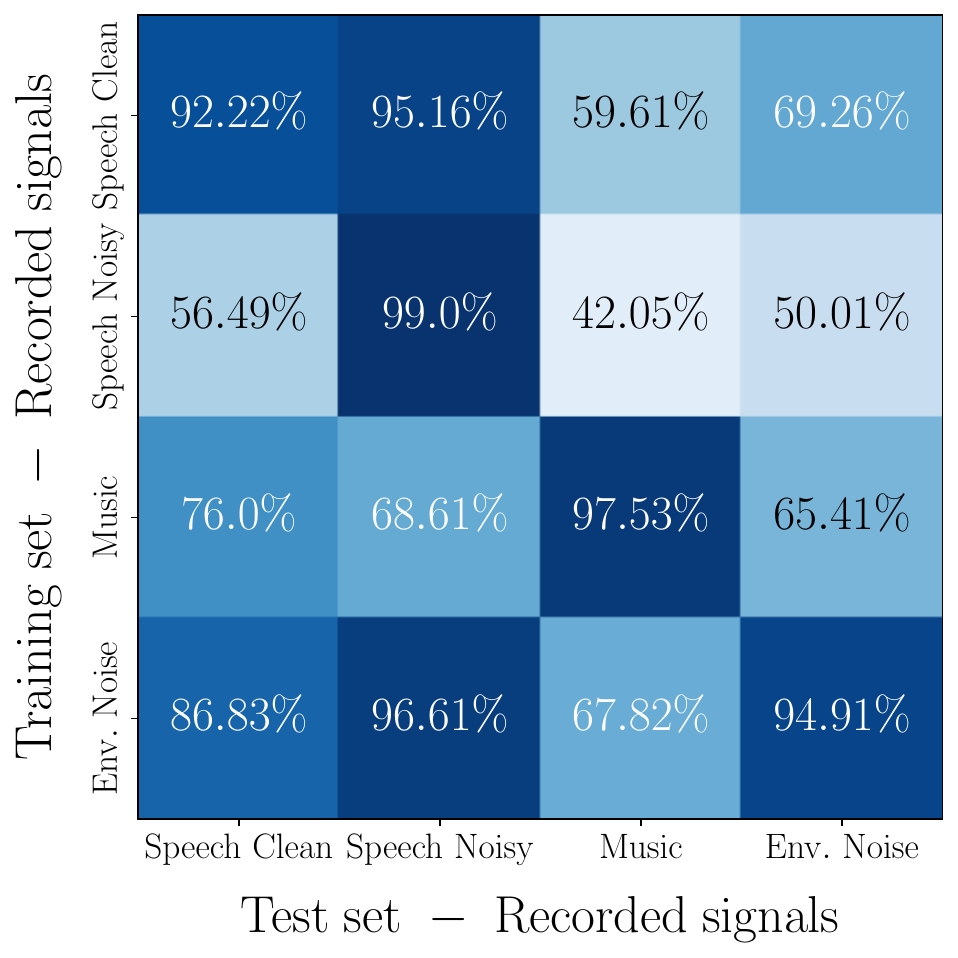}
    \caption{Balanced accuracy values of the considered classifier when trained and tested on recorded signals of different audio domains.}
    \label{fig:heatmap_original}
\end{figure}

\cref{fig:heatmap_original} shows the balanced accuracy results of this analysis, where all the training and test combinations between audio domains are explored. 
We observe that the values along the diagonal of the heatmap are consistently higher than the others, with balanced accuracy values always above \num{92}\%.
This is an expected result, as the diagonal corresponds to classifiers trained and tested on recordings from the same audio domain.

The classifier trained on clean speech exhibits strong performance on noisy speech, whereas the reverse is not observed. We hypothesize that this is due to the fact that noisy speech signals contain useful information within the entire frequency spectrum, attributable to the presence of background noise.
Consequently, the classifier trained on noisy signals may overfit on some elements present in frequency bands distinct from those of speech, resulting in diminished performance when evaluated on signals lacking these components, such as clean speech.
Our hypothesis is also supported by existing literature, which showcases methods using the microphone channel response rather than the signal content to tackle the recording smartphone model identification task~\cite{wang2011channel, cuccovillo_openset_2016}.

Regarding the considered training configurations, the classifier trained on noisy speech shows the lowest cross-test performance overall. 
Conversely, during test, the noisy speech signals are the ones that are identified best by all the classifiers. We attribute this trend to the reasons mentioned above.
In general, the best-performing classifier is the one trained on the ESC10 dataset, which contains environmental noise, with an average classification accuracy higher than \num{86}\%.

As a last experiment, we aim to assess the effectiveness of the \glspl{ir} provided in the POLIPHONE dataset.
When convolved with a clean signal, the \glspl{ir} can make it sound as if it were recorded by the microphone to which the response belongs.
From a forensic view-point and for the recording smartphone model identification task, this holds significant importance, as it enables the creation of a dataset comprising a potentially infinite number of recorded audio signals.
To evaluate this aspect, we repeat the previous analysis, training the model on the convolved tracks instead of the originally recorded ones.

\begin{figure}
    \centering
    \includegraphics[width=.8\columnwidth]{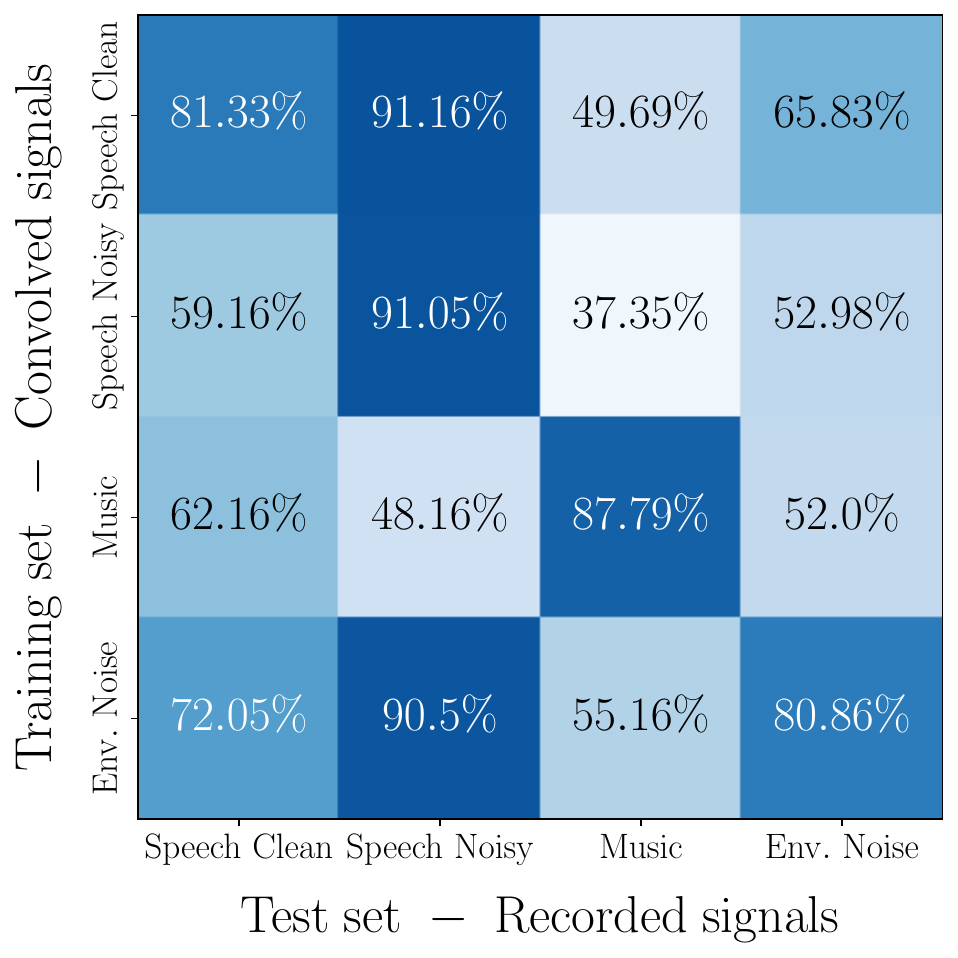}
    \caption{Balanced accuracy values of the considered classifier when trained on the convolved signals and tested on the recorded ones, in different audio domains.}
    \label{fig:heatmap_convolved}
\end{figure}

\cref{fig:heatmap_convolved} shows the results of this analysis.
The measured accuracy values exhibit an average decrease of \num{9}\% compared to those of \cref{fig:heatmap_original}.
Nonetheless, the color pattern of the two figures remains consistent: noisy speech is in general the easiest to classify (see second column), and training on environmental noise provides the best overall results (see last row).
These results validate our hypothesis: while the convolved results may not match the efficacy of the actual recordings, they enable the training of an effective classifier by considering only the \glspl{ir} of the models.
This opens the door for numerous considerations and new training strategies that can be considered to increase the robustness of the proposed classifiers.

%% file: 07_conclusion.tex
\section{Conclusion}
\label{sec:conclusion}
\glsreset{ir}

In this work we presented POLIPHONE, a dataset for the task of microphone model identification in the case of smartphones, that includes audio recordings acquired using \num{20} different recent smartphones, for a total amount of almost \num{50} hours of audio signals.
The released dataset can be useful for pushing research in the considered task and, thanks to its variety, can be employed for various types of investigations in the forensics field at large.
This corpus addresses a gap in the current state-of-the-art, where publicly available sets for the task at hand are few and outdated. 

To summarize, we bring the following contributions:
\begin{itemize}[leftmargin=*]
\item We released POLIPHONE, a comprehensive dataset designed for smartphone model identification from audio recordings, featuring audio tracks from recent devices.
\item We designed the recording setup of this dataset so that it can be expanded in the future, allowing for the inclusion of additional models and various types of audio data.
\item We benchmarked the dataset using a state-of-the-art baseline, showing the effectiveness of high-frequency content in improving classification accuracy.
\item We demonstrated the impact of training data volume on model performance, highlighting the need for extensive datasets to achieve optimal results.
\item We explored the utility of \glspl{ir} for generating training datasets in scenarios where recordings from target models are unavailable.
\end{itemize}

This is the first version of the dataset. We plan to release future developments, including more recent smartphones and different types of audio data that can be used for diverse analyses. Additionally, we will include multiple devices from the same acquisition model, enabling intra-model classification and more in-depth studies.